\documentclass{nature}

\usepackage{times}
\usepackage[dvips]{graphicx}
\usepackage{amssymb,amsfonts,amsmath}

\newcommand{\panela}{a}
\newcommand{\panelb}{b}
\newcommand{\panelc}{c}
\newcommand{\paneld}{d}

\newcommand{\panelf}{f}

\newcommand{\Methods}{\textit{Methods}}
\newcommand{\methodssection}{see \Methods}
\newcommand{\figref}[1]{Fig.~#1}
\newcommand{\figsref}[1]{Figs.~#1}
\newcommand{\eqnref}[1]{Eq.~(#1)}

\newcommand{\fignaspublicationsk}{1}
\newcommand{\fighaveshavenots}{2}
\newcommand{\fignullmodel}{3}
\newcommand{\figkctzctzp}{4}

\newcommand{\sidata}{\methodssection}
\newcommand{\sipkmixde}{\methodssection}
\newcommand{\sipkothermodels}{see \textit{Supplementary Discussion}}
\newcommand{\sipvalmatrixmcht}{see \textit{Supplementary Discussion}}

\newcommand{\sifigexpdelparams}{S1}
\newcommand{\sifigpkckp}{S2}
\newcommand{\sifigpvalmatrix}{S3}
\newcommand{\sifigensII}{S4}

\newcommand{\pvalue}{\emph{P}-value}
\newcommand{\pvalues}{\emph{P}-values}
\newcommand{\zscore}{$z$-score}

\newcommand{\nmath}{114,666}         
\newcommand{\ngc}{90,211}            
\newcommand{\nchildsints}{7,259}     
\newcommand{\nchildwithpub}{4,447}
\newcommand{\nchildnas}{269}
\newcommand{\pvalthreshold}{0.05}
\newcommand{\pvalpercentage}{5\%}
\newcommand{\nmc}{1,000}

\newcommand{\prob}[1]{p\!\left(#1\right)}

\newcommand{\probsub}[2]{p_{#1}\!\left(#2\right)}

\newcommand{\condprob}[2]{\prob{#1 | #2}}

\newcommand{\childsub}{c}
\newcommand{\parentsub}{p}
\newcommand{\degree}{k}
\newcommand{\kc}{\degree_{\childsub}}
\newcommand{\kp}{\degree_{\parentsub}}
\newcommand{\ttt}{t}
\newcommand{\tz}{\ttt}
\newcommand{\tzc}{\ttt_{\childsub}}
\newcommand{\tzp}{\ttt_{\parentsub}}

\newcommand{\pktz}{\condprob{\degree}{\tz}}

\newcommand{\pkctzc}{\condprob{\kc}{\tzc}}

\newcommand{\haves}{h}
\newcommand{\havenots}{hn}
\newcommand{\qh}{\xi_{\haves}}
\newcommand{\qhn}{\xi_{\havenots}}
\newcommand{\pih}{\pi_{\haves}}
\newcommand{\pihn}{(1-\pih)}
\newcommand{\kavemodel}{\kappa}
\newcommand{\avekavemodel}{\overline{\kavemodel}}
\newcommand{\pkde}[1]{e^{-\degree/#1}(1-e^{-1/#1})}
\newcommand{\kavehn}{\kavemodel_{\havenots}}
\newcommand{\kaveh}{\kavemodel_{\haves}}
\newcommand{\avekavehn}{\avekavemodel_{\havenots}}
\newcommand{\avekaveh}{\avekavemodel_{\haves}}
\newcommand{\bftheta}{\boldsymbol{\theta}}
\newcommand{\bfthetatz}{\bftheta_{\tz}}
\newcommand{\bfthetatzhat}{\widehat{\bftheta}_{\tz}}
\newcommand{\paramslist}{\pih,\kaveh,\kavehn}
\newcommand{\paramsset}{\{\paramslist\}}
\newcommand{\pkhavenots}{\condprob{\degree}{\kavehn}}
\newcommand{\pkhaves}{\condprob{\degree}{\kaveh}}
\newcommand{\pktztheta}{\condprob{\degree}{\bftheta}}

\newcommand{\syn}{s}
\newcommand{\Kc}{K_{\childsub}}

\newcommand{\Kcsubset}{K_{\childsub}^{\ast}}
\newcommand{\teststat}{\mathcal{S}}
\newcommand{\synteststat}{\teststat_{\syn}}

\newcommand{\kcave}{\langle\kc\rangle}
\newcommand{\kcavesyn}{\langle\kc\rangle_{\syn}}
\newcommand{\psktz}{\probsub{\syn}{\degree}}
\newcommand{\bfthetahats}{\widehat{\bftheta}_{\syn}}

\title{
The role of mentorship on prot\'eg\'e performance
}

\author{R.~Dean Malmgren$^{1,2}$, Julio M.~Ottino$^{1,3}$ \& Lu\'is A.~Nunes Amaral$^{1,3,4}$}

\begin{document}

\maketitle

\begin{affiliations}
  \item 
Department of Chemical and Biological Engineering,
Northwestern University, Evanston, IL 60208, USA

  \item 
Datascope Analytics, 
Evanston, IL 60201, USA

  \item 
Northwestern Institute on Complex Systems, 
Northwestern University, Evanston, IL 60208, USA

  \item 
Howard Hughes Medical Institute, 
Northwestern University, Evanston, IL 60208, USA

\end{affiliations}

\begin{abstract}
The role of mentorship on prot\'eg\'e performance is a matter of 
importance to academic, business, and governmental organizations.  While the
benefits of mentorship for prot\'eg\'es, mentors and their organizations are
apparent
\cite{kram85,chao92,scandura92,aryee96,allen97a,donaldson00,payne05,kram85a,higgins01},
the extent to which prot\'eg\'es mimic their mentors' career choices and
acquire their mentorship skills is
unclear \cite{allen97,green95,singh09,allen00,allen04,ragins99,paglis06}.
Here, we investigate one aspect of mentor emulation by studying mentorship
fecundity---the number of prot\'eg\'es a mentor trains---with data from the
Mathematics Genealogy Project \cite{mathgenealogy07}, which tracks the
mentorship record of thousands of mathematicians over several centuries.  We
demonstrate that fecundity among academic mathematicians is correlated with
other measures of academic success.  We also find that
the average fecundity of mentors remains stable over 60 years of recorded
mentorship.
We further uncover three significant correlations in mentorship fecundity.
First, mentors with small mentorship fecundity train prot\'eg\'es that go on to
have a 37\% larger than expected mentorship fecundity.  Second, in the first
third of their career, mentors with large fecundity train prot\'eg\'es that go
on to have a 29\% larger than expected fecundity.  Finally, in the last third
of their career, mentors with large fecundity train prot\'eg\'es that go on to
have a 31\% smaller than expected fecundity.

\end{abstract}


A large body of literature supports the hypothesis that prot\'eg\'es and
mentors benefit from the mentoring relationship \cite{kram85,chao92}.
Prot\'eg\'es that receive career coaching and social support, for instance, are
reportedly more likely to have high performance ratings, a higher salary, and
receive promotions \cite{kram85,scandura92}.  In return, mentors receive
fulfillment not only by altruistically improving the welfare of their
prot\'eg\'es, but also by improving their own
welfare \cite{aryee96,allen97a,allen97}.  Organizations benefit as well, since
prot\'eg\'es are more likely to be committed to their
organization \cite{donaldson00,payne05} and exhibit organizational citizenship
behavior \cite{donaldson00}.  These benefits are not only obtained through the
traditional dyadic mentor-prot\'eg\'e relationship, but also through peer
relationships that supplement prot\'eg\'e development \cite{kram85a,higgins01}.

The benefits of mentorship underscore the importance of understanding how
mentors were in turn trained to foster the development of outstanding mentors.
One might suspect that prot\'eg\'es learn managerial approaches and
motivational techniques from their mentors and, as a result, emulate their
mentorship methodologies; this suggests that outstanding mentors are trained by
other outstanding mentors.  This possibility is sometimes formalized as the
rising star hypothesis \cite{green95,singh09}; it postulates that mentors
select up-and-coming prot\'eg\'es based on their perceived ability, potential
and past performance \cite{allen97,allen00,allen04}, including promotion
history and proactive career behaviors \cite{singh09}.  Rising-star
prot\'eg\'es are reportedly more likely to ``intend to mentor'', resulting in a
``perpetual cycle'' of rising-star prot\'eg\'es that emulate their mentors by
seeking other rising stars as their prot\'eg\'es \cite{ragins99}.

However, there is conflicting evidence concerning the rising star
hypothesis \cite{paglis06}, so the extent to which prot\'eg\'es
mimic their mentors remains an open question.  Indeed, we are unaware of any
studies that systematically track mentorship success over the entire career of
a mentor, so the validity of the rising star hypothesis has yet to be fully
explored.  
Here, we investigate whether prot\'eg\'es acquire the mentorship skills of
their mentors by studying mentorship fecundity---the number of prot\'eg\'es
that a mentor trains over the course of their career.
This measure is advantageous as it directly measures an outcome of the
mentorship process that is relevant to sustained mentorship, allowing us to
quantify the degree to which mentor fecundity determines prot\'eg\'e fecundity.


Scientific mentorship offers a unique opportunity for studying this question
because there is a structured mentorship environment between advisor and
student that is, in principle, readily accessible \cite{bourne08,enserink09}.
We study a prototypical mentorship network collected from the Mathematics
Genealogy Project \cite{mathgenealogy07}, which aggregates the graduation date,
mentor, and prot\'eg\'es of \nmath~mathematicians from as early as 1637.  From
this information, we construct a network where links are formed from a mentor
to each of his $\degree$ prot\'eg\'es, where $\degree$ denotes mentorship
fecundity.  This database is unique because it explicitly tracks the
career-long mentorship record of a large population of mentors within a single
discipline.  We focus here on the \nchildsints~mathematicians that graduated
between 1900 and 1960 since their mentorship record is the most reliable
(\sidata).

Although the mentorship records gathered from the Mathematics Genealogy Project
provide the most comprehensive data source available for studying academic
performance throughout a mathematician's career, there are obviously other
plausible metrics for evaluating academic
performance \cite{king87,moed05b,hirsch05}.  We have also compared the
mentorship data against a list of publications
for \nchildwithpub~mathematicians and a list of \nchildnas~inductees into the
United States' National Academy of Sciences (NAS)
(\sidata).  We find that mentorship fecundity is much
larger for NAS members than for non-NAS members
(\figref{\fignaspublicationsk\panela}).  We further find that the
number of publications is strongly correlated with fecundity, regardless of
whether or not a mathematician is a NAS member
(\figref{\fignaspublicationsk\panelb}).  These results demonstrate
that, although fecundity is not a typical measure of academic performance, it
is closely related to other measures of academic success.  Thus, even though
our investigation concerns how fecundity is correlated between mentor and
prot\'eg\'e, our results also address questions in the academic evaluation
literature concerning the success of a mathematician.


We first investigate whether it is possible to predict the fecundity of a
mathematician by modeling the fecundity distribution $\pktz$ as a function of
graduation year $t$.  Considering that some mathematicians remain in academia
throughout their career while others spend only a portion of their career in
academia, one might expect that there are two types of individuals when it
comes to academic mentorship fecundity---``haves'' and ``have-nots''---in the
sense that these mathematicians have or have not had the opportunity to mentor
students throughout their career.  If each mentor chooses to train a new
academic prot\'eg\'e with probability $\qh$ or $\qhn$ and stops training
academic prot\'eg\'es otherwise, depending on whether they are a ``have'' or
``have-not'' respectively, then we would expect that the resulting fecundity
distribution is a mixture of two discrete exponential distributions
\begin{equation}
\pktztheta = \pih\pkhaves+\pihn\pkhavenots ,
\label{eqn:pktztheta}
\end{equation}
where $\pih$ is the probability that a mathematician is a ``have'', and
$\pkhaves$ and $\pkhavenots$ are discrete exponential distributions
$\condprob{\degree}{\kavemodel}=\pkde{\kavemodel}$ with average fecundity
$\kaveh=1/\ln\qh^{-1}$ and $\kavehn=1/\ln\qhn^{-1}$ for ``haves'' and
``have-nots'' respectively.  We estimate the parameters $\bftheta=\paramsset$
of this distribution from the empirical data using
expectation-maximization \cite{bishop07}.  Using Monte Carlo hypothesis testing
(\sipkmixde), we have found that
\eqnref{\ref{eqn:pktztheta}}~can not be rejected as a candidate description of
the fecundity distribution $\pktz$.  For an alternative description of the
fecundity distribution
$\pktz$, \sipkothermodels~and \figref{\sifigexpdelparams}.

As one might expect, the probability $\pih$ that an individual is a ``have''
experiences dramatic changes over time due to historical events, such as two
World Wars, the beginning of the Cold War, and considerable increases in
academic funding
(\figref{\fighaveshavenots}\panelb).  In contrast, the average
fecundities of ``haves'' and ``have-nots'' \emph{do not} exhibit systematic
historical changes---$\avekaveh=9.8\pm0.4$ and
$\avekavehn=0.47\pm0.03$---suggesting that these quantities offer fundamental
insight into the mentorship process among mathematicians
(\figref{\fighaveshavenots}\panelc--\paneld).

The stationarity of $\kaveh$ and $\kavehn$ also provides a simple heuristic for
classifying an individual as a ``have'' or a ``have-not''; by maximum
likelihood, an individual is a ``have'' if $k\ge2$ and a ``have-not''
otherwise.  These results raise the possibility that similar features, perhaps
with different characteristic scales of fecundity, may be present in other
mentorship domains.


While our description of the fecundity distribution has highlighted a
fundamental property of mentorship among mathematicians, it is not predictive
of the behavior of individual mathematicians in the sense that fecundity,
according to this model, is a random variable drawn from a distribution
of \eqnref{\ref{eqn:pktztheta}}.  We next test whether prot\'eg\'es mimic the
mentorship fecundity of their mentors by comparing prot\'eg\'e fecundity with a
suitable null model that does not introduce correlations in fecundity.  In
analogy with ancestral genealogies and their notion of parents giving birth to
children, networks generated from uncorrelated branching processes offer a
useful and appropriate context for studying the mathematician genealogy
network.  Here, a graduation date is equivalent to a birth date and mentors and
prot\'eg\'es are equivalent to parents and children, respectively.  We will
consequently use the subscripts $\parentsub$ and $\childsub$ when it is
necessary to make generational statements relating parents and children.

In a branching process \cite{athreya04}, a parent $\parentsub$, born at time
$\tzp$, has $\kp$ children.  A child $\childsub$ of parent $\parentsub$ is born
at time $\tzc$ and subsequently has $\kc$ children.  The fecundity $\degree$ of
each individual is drawn from the conditional fecundity distribution $\pktz$
for an individual born at time $\tz$.  Networks generated from this type of
branching process are therefore defined by the birth date of each individual
$\tz$, the fecundity distribution $\pktz$, and the chronology of child births
$\{\tzc\}$ for each parent (\figref{\fignullmodel\panela}).

We compare the mathematician genealogy network with two ensembles of randomized
genealogies from the branching process family.
Random networks from Ensemble I retain the birth date of each individual
$\tz$, the fecundity $\degree$ of each individual, and the chronology of child
births $\{\tzc\}$ for each parent (\figref{\fignullmodel\panelb}).
Random networks from Ensemble II additionally restrict parent--child pairs to
have the same age difference ($\tzc-\tzp$) as parent--child pairs in the
empirical network (\figref{\fignullmodel\panelc}).
All other attributes of these networks are randomized using a link switching
algorithm (\methodssection) \cite{milo03,itzkovitz04}, so neither of these random
network ensembles introduces correlations between parent fecundity and child
fecundity or temporal correlations in fecundity, providing a suitable basis for
comparison with the mathematician genealogy network.


To explore the influence of mentor fecundity and age difference on
prot\'eg\'e fecundity, we partition prot\'eg\'es according to the
fecundity of their mentors and the age difference between mentor and
prot\'eg\'e ($\tzc-\tzp$).
Given our findings (\sipvalmatrixmcht, \figsref{\sifigpkckp--\sifigpvalmatrix}),
it is clear that age differences impact fecundity in a non-random manner for
prot\'eg\'es whose mentors have $\kp<3$.  We partition the remaining
prot\'eg\'es whose mentors have $\kp\ge3$ into two groups: prot\'eg\'es whose
mentors are below-average ``haves'' ($3\le\kp<10$) and prot\'eg\'es whose
mentors are above-average ``haves'' ($\kp\ge10$).  We then partition these
three groups of prot\'eg\'es according to when they graduated during their
mentors' career.  Specifically, we split each group of prot\'eg\'es into
terciles, the most fine-grained grouping that still gives us sufficient power
to examine the statistical significance of any differences between the
empirical data and the null models.

We use the partitioning of children into classes to examine the relationship
between the average child fecundity $\kcave$ and the age difference $\tzc-\tzp$
between parent and child
(\figsref{\figkctzctzp\panela,\panelb~and \sifigensII\panela,\panelb}).
If the data are consistent with a branching process, then we would expect the
average child fecundity $\kcave$ to exhibit no temporal dependence.  However,
the regressions between the average child fecundity \zscore~(\methodssection) and the
age difference between parent and child $\tzc-\tzp$ deviate significantly
(\figsref{\figkctzctzp\panelc~and \sifigensII\panelc}) from this
expectation for both random ensembles to reveal three distinct features.
First, mentors with $\kp<3$ train prot\'eg\'es that go on to have a 37\% larger
than expected mentorship fecundity throughout their career.  Second, in the
first third of their career, mentors with $\kp\ge10$ train prot\'eg\'es that go
on to have a 29\% larger than expected fecundity.  Finally, in the last third
of their career, mentors with $\kp\ge10$ train prot\'eg\'es that go on to have
a 31\% smaller than expected fecundity.

The fact that mentors with $\degree<3$ train prot\'eg\'es with larger than
expected fecundity throughout their career is somewhat counter-intuitive.
According to the rising star hypothesis~\cite{green95,singh09}, one might have
expected that prot\'eg\'es trained by mentors with $\degree<3$ are likely to
mimic their mentors and therefore have smaller than expected fecundity.  Our
results demonstrate that this is not the case.  One possible explanation is
that mentors with $\degree<3$ are more aware of the resources they must
allocate for effective mentorship, leading to a more enriching mentorship
experience for their prot\'eg\'es.  An alternative hypothesis is that mentors
with $\degree<3$ select for, or are selected by, prot\'eg\'es that have a
greater aptitude for mentorship.

The striking temporal correlations for mentors with $\kp\ge10$ are intriguing
as well.  Since mentors with $\kp\ge10$ represent the upper echelon of mentors
in mathematics, these mentors are likely ``rising stars'' early in their
academic career.  The fact that these mentors train prot\'eg\'es with large
fecundity early in their career supports the rising star hypothesis.

By the end of these mentor's careers, however, their prot\'eg\'es have smaller
than expected fecundity.  Perhaps mentors, who ultimately have large fecundity,
spend less and less resources training each of their prot\'eg\'es as their
career progresses.  Alternatively, prot\'eg\'es with large mentorship fecundity
aspirations might court prolific mentors early in their mentor's career whereas
prot\'eg\'es with small fecundity aspirations might court prolific mentors
later in their mentor's career.  Our findings therefore reveal interesting
nuances to the rising star hypothesis.

It is unclear whether the temporal correlations we uncover in mentorship
fecundity might generalize beyond mathematicians in academia.  Anecdotally,
mathematicians are thought to perform their best work at a young
age~\cite{hardy40}, a perception that may influence how mentors and
prot\'eg\'es choose each other.  Perceptions in other domains, however, may
differ and subsequently influence mentor and prot\'eg\'e selection in different
ways.  As data for other academic disciplines \cite{bourne08,enserink09},
business and the government becomes available, it will be important to
determine whether temporal correlations in fecundity are a general consequence
of mentorship, or a particular consequence of mentorship for mathematicians in
academia.

Regardless, our results offer another means of judging academic impact in
science as well as the impact of managers on their employees, both of which are
notoriously complicated and risky affairs.  These assessments are
multi-dimensional, metrics and expectations are domain dependent, and placement
of creative output, time-scales of impact and recognition vary significantly
from field to field.  Ultimately, assessment of individuals for awards and
promotion is based on painstaking individual analysis by selection committees
and peers.  While these committees may have varying goals and incentives, it is
important that collective arguments---the kind of arguments we are making
here---be based on sound quantitative analysis.  Although the extent to which
our findings extrapolate to other domains may vary, we are confident that the
kind of analysis presented here will serve to elevate the discourse on
scientific and managerial impact.

\begin{methodssummaryenv}

{\noindent \bf Data acquisition.}
We use data from the Mathematics Genealogy Project \cite{mathgenealogy07} to
identify the \nchildsints~prot\'eg\'e mathematicians that are in the giant
component \cite{stauffer92} and graduated between 1900 and 1960, of
which \nchildwithpub~of them have linked publication records through
MathSciNet.  We use a text matching algorithm \cite{chapman00} to
semi-automatically match members of the National Academy of Science with
mathematicians from the Mathematics Genealogy Project.

{\noindent \bf Monte Carlo hypothesis testing for $\boldsymbol{\pktz}$.}
We use Monte Carlo hypothesis testing~\cite{dagostino86} to determine whether
\eqnref{\ref{eqn:pktztheta}}~with maximum-likelihood~\cite{bishop07} parameters
$\bftheta$ can be rejected as a candidate model for $\pktz$ at the
$\alpha=\pvalthreshold$ significance level.

{\noindent \bf Random network generation}. 
We use a variation of the Markov chain Monte Carlo
algorithm \cite{milo03,itzkovitz04} to construct each of the \nmc~random
networks in Ensembles I and II.  Specifically, we restrict the switching of
endpoints of links $\parentsub\rightarrow\childsub$ that belong to the same
link class $\mathcal{L}$, where the link classes are defined as
$\mathcal{L}_{\rm I}(t)=\{\parentsub\rightarrow\childsub|\tzc=t\}$ and
$\mathcal{L}_{\rm II}(s,t)=\{\parentsub\rightarrow\childsub|\tzp=s,\tzc=t\}$
for networks from Ensembles I and II, respectively.  Each link class
$\mathcal{L}$ can be thought of as a subgraph, which can then be randomized in
the usual way by attempting 100 switches per link in each link class
$\mathcal{L}$ \cite{milo03,itzkovitz04}.

{\noindent \bf Average fecundity z-score.}
By the central limit theorem, the average of variates drawn from $\pkctzc$ is
normally distributed since $\pkctzc$ is well-described by a mixture of discrete
exponential distributions, a distribution with finite variance.  Given a set of
child fecundities $\Kc=\{\kc\}$, we quantify how significantly a subset of
these child fecundities $\Kcsubset\subset\Kc$ deviates from $\Kc$ by measuring
the \zscore~of the average child fecundity $\kcave$ of all nodes within the
subset $\Kcsubset$ compared with the average child fecundity $\kcavesyn$
computed from children within an equivalent subset $\Kcsubset$ in the synthetic
networks.  That is, we compute $z=(\kcave-\mu)/\sigma$ where $\mu$ is the
ensemble average of $\{\kcavesyn\}$ and $\sigma$ is the standard deviation of
the ensemble $\{\kcavesyn\}$ over the \nmc~realizations generated for our null
models.

\end{methodssummaryenv}

\vspace*{1cm}

\begin{addendum}
\item[Supplementary Information] is linked to the online version of the paper
  at www.nature.com/nature.
\item 
We thank R.~Guimer\`a, P.~McMullen, A.~Pah, M.~Sales-Pardo, E.N.~Sawardecker,
D.B.~Stouffer, and M.J.~Stringer for insightful comments and suggestions.
L.A.N.A. gratefully acknowledges the support of NSF awards SBE 0830388 and IIS
0838564. 
All figures were generated with PyGrace (http://pygrace.sourceforge.net) with
color schemes from http://colorbrewer.org.


\item[Author Contributions] R.D.M. analyzed data, designed the study, and wrote
  the paper.  J.M.O.~and L.A.N.A.~designed the study and wrote the paper.
\item[Author Information] Reprints and permissions information is available at
  npg.nature.com/reprintsandpermissions.  The authors declare that they have no
  competing financial interests.  Correspondence and requests for materials
  should be addressed to J.M.O. (jm-ottino@northwestern.edu) or
  L.A.N.A.~(amaral@northwestern.edu).
\end{addendum}

\clearpage
\begin{methodsenv}
\section*{Mathematics Genealogy Project data}

We study a prototypical mentorship network collected from the Mathematics
Genealogy Project \cite{mathgenealogy07}, which aggregates the graduation date,
mentor, and advisees of \nmath~mathematicians from as early as 1637.  From this
information, we construct a mathematician genealogy network where links are
formed from a mentor to each of his $\degree$ prot\'eg\'es.

The data collected by the Mathematics Genealogy Project are self-reported, so
there is no guarantee that the observed genealogy network is a complete
description of the mentorship network.  In fact, 16,147 mathematicians do not
have a recorded mentor and, of these, 8,336 do not have any recorded
prot\'eg\'es.  To avoid having these mathematicians distort our analysis, we
restrict our analysis to the \ngc~mathematicians that comprise the giant
component~\cite{stauffer92} of the network;
that is, we restrict our analysis to the largest set of connected
mathematicians in the mathematician genealogy network.

Although the Mathematics Genealogy Project contains information on
mathematicians from as early as 1637, this does not necessarily indicate that
all of these records are representative of the evolution of the network.  For
example, prior to 1900, the Project records fewer than 52 new graduates per
year worldwide.  Furthermore, since mathematicians oftentimes have mentorship
careers lasting 50 years or more (\figref{\ref{fig:graduation_rate}}), we are
not guaranteed to have complete mentorship records for mathematicians that
graduated after 1960.  We therefore restrict our analysis to
the \nchildsints~prot\'eg\'e mathematicians that graduated between 1900 and
1960, for whom we believe that the graduation and mentorship record is the most
reliable.

\section*{MathSciNet data}

Of the \nchildsints~prot\'eg\'e mathematicians that graduated between 1900 and
1960, \nchildwithpub~of them have linked MathSciNet publication records which
are used in our analysis.

\section*{U.S.~National Academy of Science data}

The United States' National Academy of Science maintains two databases of its
membership.  The first database consists of all deceased members elected to the
Academy from as early as 1863.  This database records the name of the inductee,
their election year, their date of death, and a link to a biographical sketch.
The second database consists of all active members of the Academy.  This
database records the name of the inductee, their institution, their academic
field, and their election year.

The challenge to matching this data with the Mathematics Genealogy Project data
is that there is no direct link between a member of the National Academy and
Mathematics Genealogy Project page and vice versa.  This ambiguity is somewhat
confounded by the fact that some members of the Academy have common names.  To
circumvent these problems, we use a text matching algorithm~\cite{chapman00} to
semi-automatically detect if a member of the Academy matches a name in the
Mathematics Genealogy Project database.  We use this procedure to curate the
\nchildnas~members of the Academy that definitively match mathematicians in the
Mathematics Genealogy Project database.

\section*{Monte Carlo hypothesis testing for $\boldsymbol{\pktz}$}

Given a model $\mathcal{M}$ with parameters $\bfthetatz$ for the empirically
observed fecundity distribution $\pktz$, we use Monte Carlo hypothesis testing
to determine whether the model $\mathcal{M}$ can be rejected as a candidate
model for $\pktz$~\cite{dagostino86}.  The Monte Carlo hypothesis testing
procedure is as follows.
First, we calculate the best-estimate parameters $\bfthetatzhat$ for
model $\mathcal{M}$ at time $\tz$ using maximum likelihood
estimation~\cite{bishop07}.  Second, we compute the test statistic $\teststat$
(detailed below) between the model $\mathcal{M}(\bfthetatzhat)$ and the
empirical fecundity distribution $\pktz$.
We next generate a synthetic fecundity distribution $\psktz$ from model
$\mathcal{M}(\bfthetatzhat)$ using the best-estimate parameters
$\bfthetatzhat$, and we treat the synthetic data exactly the same as we treated
the empirical data: first, we calculate the best-estimate parameters
$\bfthetahats$ for model $\mathcal{M}$ from maximum likelihood estimation;
second, we compute the test statistic $\synteststat$ between the model
$\mathcal{M}(\bfthetahats)$ and the synthetic fecundity distribution $\psktz$.
We generate synthetic fecundity distributions $\psktz$ and their corresponding
synthetic test statistics $\synteststat$ until we accumulate an ensemble of
1,000 Monte Carlo test statistics $\{\synteststat\}$.  Finally, we calculate a
two-tailed \pvalue~with a precision of 0.001.
As is customary in hypothesis testing, we reject the model $\mathcal{M}$ at
time $\tz$ if the \pvalue~is less than a threshold value.  We
select a \pvalue~threshold of \pvalthreshold; that is, if less than
\pvalpercentage~of the synthetic data sets exhibit deviations in the test
statistic that are larger than those observed empirically, the model is
rejected at time $\tz$.

Since we are conducting hypothesis tests with the fecundity distribution
$\pktz$---a distribution with a discrete support---it is important to use a
test statistic $\teststat$ that is appropriate for testing discrete
distributions.  We use the $\chi^2$ test statistic where we bin $\pktz$ such
that each bin has at least one expected observation according to the model
$\mathcal{M}(\bfthetatzhat)$.  This binning prevents observations that are
exceptionally rare from dominating our statistical test and skewing our
results.

\section*{Random network generation} 

We use Markov chain Monte Carlo algorithm \cite{milo03,itzkovitz04} to build
random networks from the mathematician genealogy network.  The standard version
of this algorithm inherently preserves the fecundity of each individual, but it
does not preserve the chronology of child births $\{\tzc\}$ for each parent.
To obtain random networks belonging to Ensemble I or Ensemble II, we restrict
the switching of endpoints of links $\parentsub\rightarrow\childsub$ that
belong to the same link class $\mathcal{L}$, where the link classes are defined
as $\mathcal{L}_{\rm I}(t)=\{\parentsub\rightarrow\childsub|\tzc=t\}$ and
$\mathcal{L}_{\rm II}(s,t)=\{\parentsub\rightarrow\childsub|\tzp=s,\tzc=t\}$
for networks from Ensembles I and II, respectively.  Each link class
$\mathcal{L}$ can be thought of as a subgraph, which can then be randomized
using the Markov chain Monte Carlo algorithm.  Here, we attempt 100 switches
per link in each link class $\mathcal{L}$, which sufficiently alters random
networks away from the original empirical network \cite{milo03,itzkovitz04}.
We repeat this procedure \nmc~times to generate a set of \nmc~random networks
for each ensemble.  

\section*{Average fecundity z-score}

The average of variates drawn from $\pkctzc$ is normally distributed since
$\pkctzc$ is well-described by a mixture of discrete exponential distributions,
a distribution with finite variance, and thus the central limit theorem
applies.  Given a set of child fecundities $\Kc=\{\kc\}$, we quantify how
significantly a subset of these child fecundities $\Kcsubset\subset\Kc$
deviates from $\Kc$ by measuring the \zscore~of the average child fecundity
$\kcave$ of all nodes within the subset $\Kcsubset$ compared with the average
child fecundity $\kcavesyn$ computed from children within an equivalent subset
$\Kcsubset$ in the synthetic networks.  That is, we compute
$z=(\kcave-\mu)/\sigma$ where $\mu$ is the ensemble average of $\{\kcavesyn\}$
and $\sigma$ is the standard deviation of the ensemble $\{\kcavesyn\}$ over the
\nmc~realizations generated for our null models.

\end{methodsenv}


\clearpage
\newcommand{\figAcaption}{
Relationship between mentorship fecundity and other performance metrics.
{\bf \panela}, Cumulative distribution of the mentorship fecundity for NAS
members (red) and non-NAS members (black).  NAS members have an average
fecundity of $\langle\degree\rangle_{\rm NAS}=14$, which is far greater than
the average fecundity of non-NAS members $\langle\degree\rangle_{\rm
  non-NAS}=3.1$, indicating that fecundity is closely related to academic
recognition.  Note that not all mathematicians in the non-NAS group were
elegible for NAS membership due to citizenship and other circumstances.  This
fact makes the result in the figure all the more striking.
{\bf \panelb}, Average and standard error (symbols and error bars) of the
number of publications as a function of the mentorship fecundity for NAS
members (red) and non-NAS members (black).  NAS members have nearly twice as
many publications on average as non-NAS members for all fecundity levels.
}

\begin{figure}
\centerline{\includegraphics[width=0.7\columnwidth]{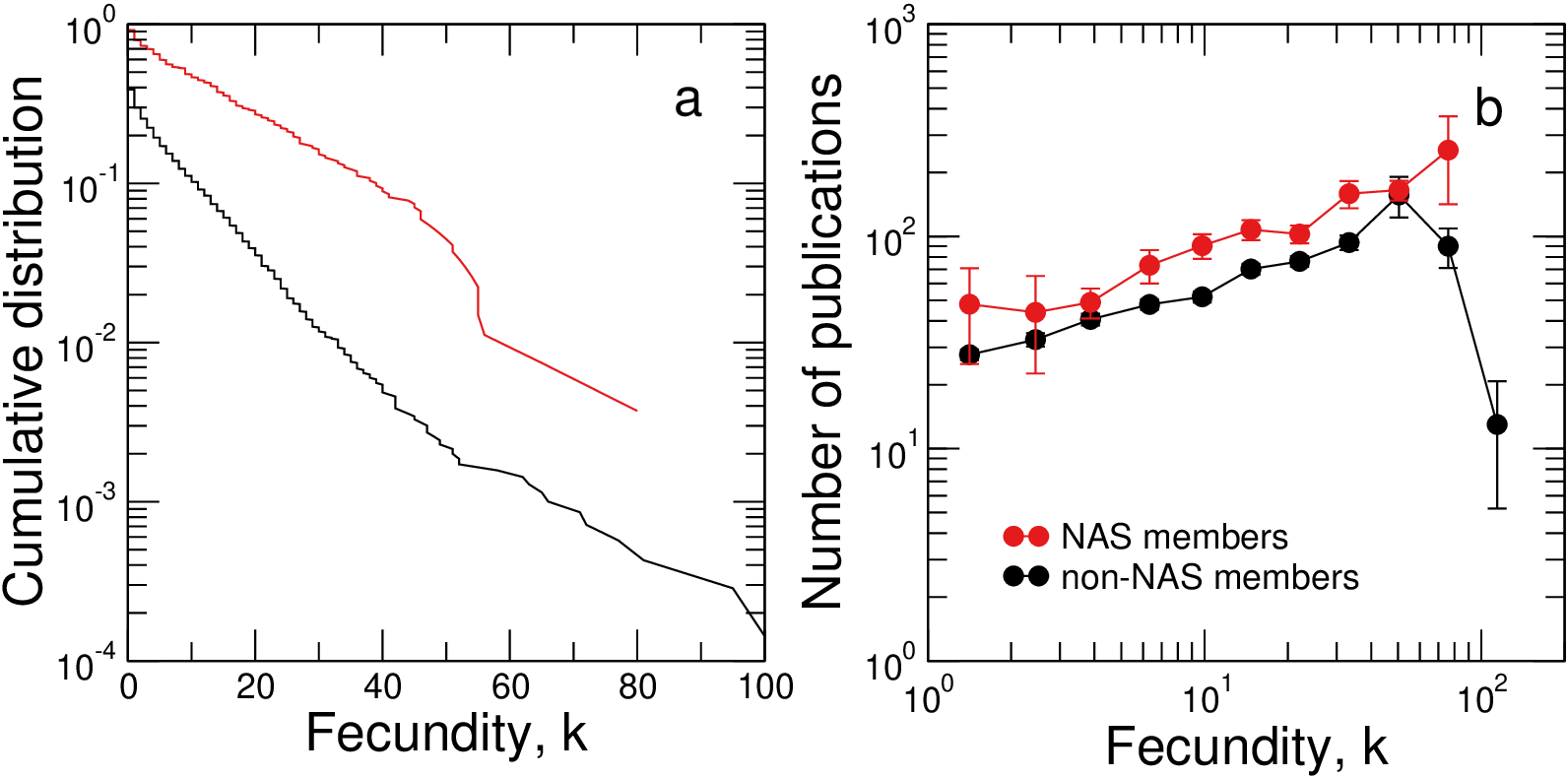}}
\caption{\figAcaption}
\label{fig:nas_publications_k}
\end{figure}


\clearpage
\newcommand{\figBcaption}{
Evolution of the fecundity distribution.
{\bf \panela}--{\bf \panelc}, Cumulative distribution of the fecundity of
mathematicians that graduated during 1910, 1930, and 1950 (symbols) compared
with the best-estimate predictions of a mixture of two discrete exponentials
(lines).  Monte Carlo hypothesis testing confirms that
this model can not be rejected as a model of the fecundity distribution during
every year from 1900--1960, as denoted by the \pvalues~above the
$\alpha=\pvalthreshold$ significance level (\sipkmixde).
{\bf \paneld}--{\bf \panelf}, Best-estimate parameters calculated by maximum
likelihood for a mixture of two discrete exponentials as a function of time.
Dashed lines denote average parameter values between 1900--1960 and colored
circles denote the years displayed in panels \panela--\panelc.  The probability
$\pih$ of being an ``have'' changes over time, generally corresponding with
historic events (hashed grey shading).  In contrast, the average fecundities
$\avekaveh=9.8\pm0.4$ and $\avekavehn=0.47\pm0.03$ remain stable until 1960, at
which point they steadily decrease (grey shaded region), corresponding with the
time at which mentorship records become incomplete
(\sidata).
}

\begin{figure}
\centerline{\includegraphics[width=1.0\columnwidth,height=0.6\textheight,keepaspectratio=true]{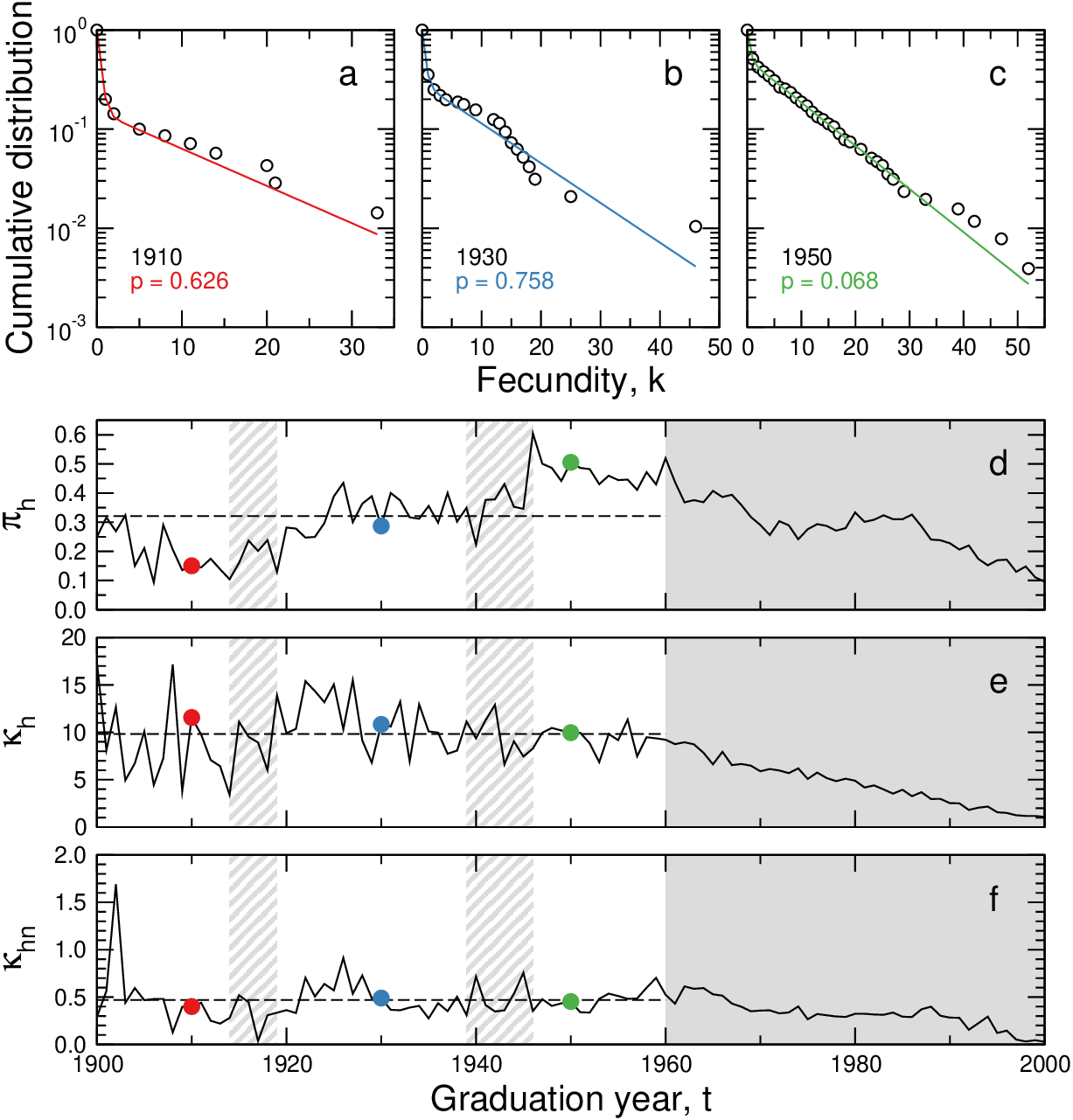}}
\caption{\figBcaption}
\label{fig:haves_have-nots}
\end{figure}


\clearpage
\newcommand{\figCcaption}{
Branching process null models.
{\bf \panela}, Subset of the mathematician genealogy network.
Mentors/parents (black circles) are connected to each of their
prot\'eg\'es/children (white circles).  The horizontal positions of
mathematicians represent their graduation/birth date $\tz$.  The bottom
two parents were born in 1924, the top two parents were born in 1937, and all
four parents have a child born in 1958.
From the parent's perspective, three essential features of the empirical
network must be preserved in random networks generated from the two branching
process null models: the birth date $\tzp$, the fecundity $\kp$, and the
chronology of child births $\{\tzc\}$.
{\bf \panelb}, Random networks from Ensemble I preserve these three essential
features.  Solid lines highlight the links in the empirical network whose end
points can be randomized.  Dashed lines illustrate one of the possible
randomization moves after switching the corresponding pair of links.  Note that
the age difference between parent and child is not preserved.
{\bf \panelc}, Random networks from Ensemble II preserve these three essential
features as well as the age difference between parent and child.  Solid lines
of the same color highlight the links in the empirical network whose end
points can be randomized.  Dashed lines illustrate one of the possible
randomization moves after switching the corresponding pair of links.
Random networks for each ensemble are generated by attempting 100 switches per
link (\methodssection).
}

\begin{figure}
\centerline{\includegraphics[width=0.33\columnwidth]{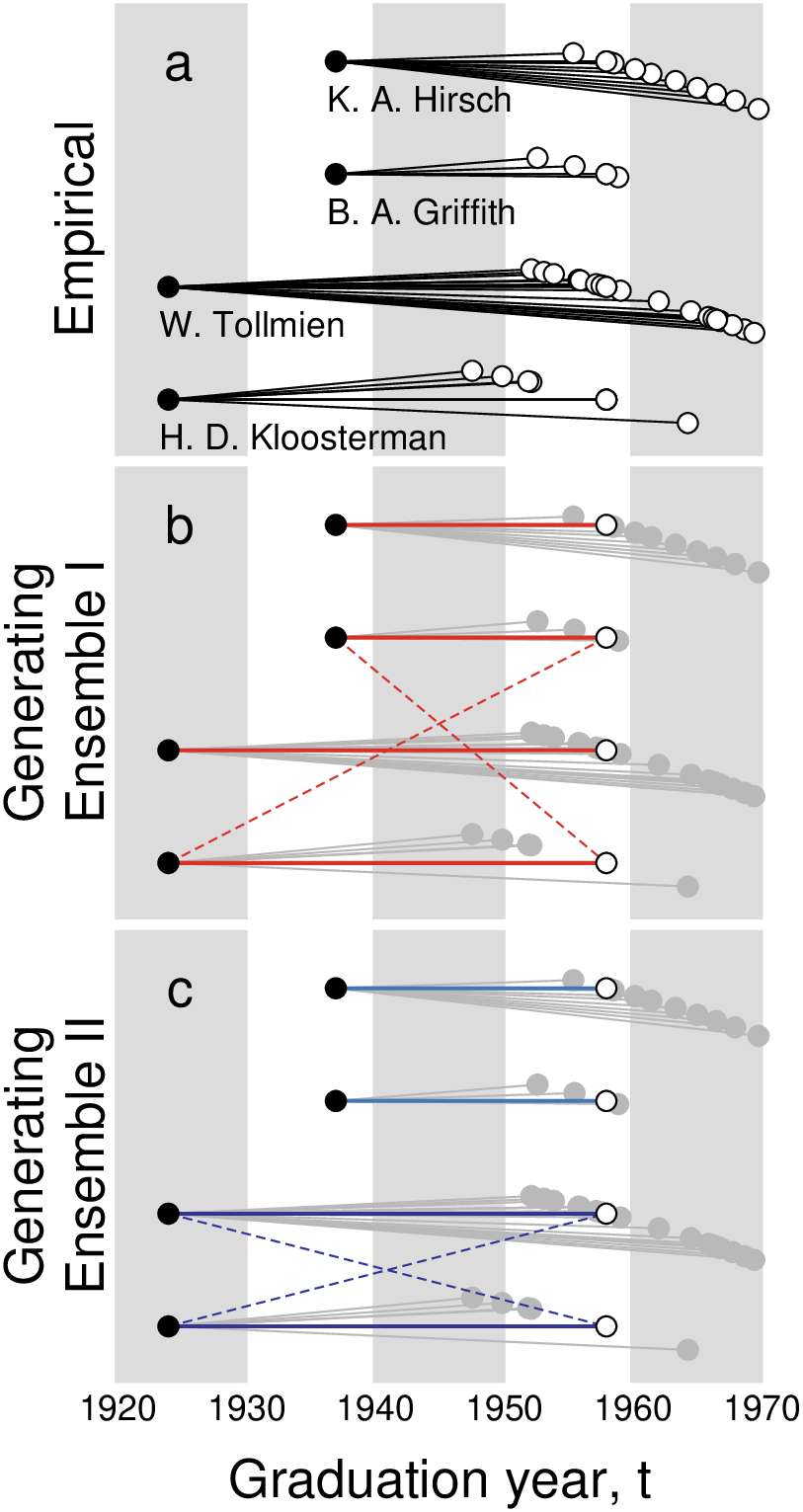}}
\caption{\figCcaption}
\label{fig:null_model}
\end{figure}


\clearpage
\newcommand{\figcapens}{I}
\newcommand{\figcapensalt}{II}
\newcommand{\figcapensaltfig}{\figref{\sifigensII}}
\newcommand{\pvala}{=0.009}
\newcommand{\pvalb}{=0.366}
\newcommand{\pvalc}{<0.001}
\newcommand{\figDcaption}{
Effect of age difference $\tzc-\tzp$ between mentor and prot\'eg\'e on
prot\'eg\'e fecundity.
{\bf \panela}, Fecundity distribution of children born during the 1910s from
parents with $\kp<3$, $3\le\kp<10$, and $\kp\ge10$ compared with the
expectation from Ensemble \figcapens~(grey line).  We separate children into
terciles (early, middle, late) according to the time difference in birth dates
$\tzc-\tzp$ between parents and children.  Note that the average fecundity of
children born from parents with $\kp<3$ is larger than expected, regardless of
whether they were born during the early, middle, or later part of their
parent's life.  Note also that the average fecundity of children born from
parents with $\kp\ge10$ decreases throughout their parent's life.
{\bf \panelb}, We quantify the significance of these trends during each decade
(colored symbols) by computing the \zscore~of the average child fecundity
$\kcave$ compared with the average child fecundity in networks from
Ensemble \figcapens.  This information is summarized by identifying the linear
regression (solid black line).  Note that the regression lines for networks
from our null model (grey lines) vary around the expectation of our null model
(dashed black line).
{\bf \panelc}, Significance of linear regressions in panel \panelb.  We compare
the slope and intercept of the empirical regression (black circle) with the
distribution of the slope and intercept of the same quantities computed from
the null model.  Since these quantities are approximately distributed as a
multivariate Gaussian, we compute the equivalent of a two-tailed \pvalue~by
finding the fraction of synthetically generated slope--intercept pairs that lie
outside of the equi-probability surface of the multivariate Gaussian (dashed
ellipse).  Note that the slope and intercept of the regression for
children from parents with small ($p\pvala$) and large fecundity ($p\pvalc$)
are significantly different from the expectation for the null model, consistent
with the data displayed in panel \panela.
Comparisons with expectations from random networks from
Ensemble \figcapensalt~yield the same conclusions (\figcapensaltfig).

%
}

\begin{figure}
\centerline{\includegraphics[width=\columnwidth]{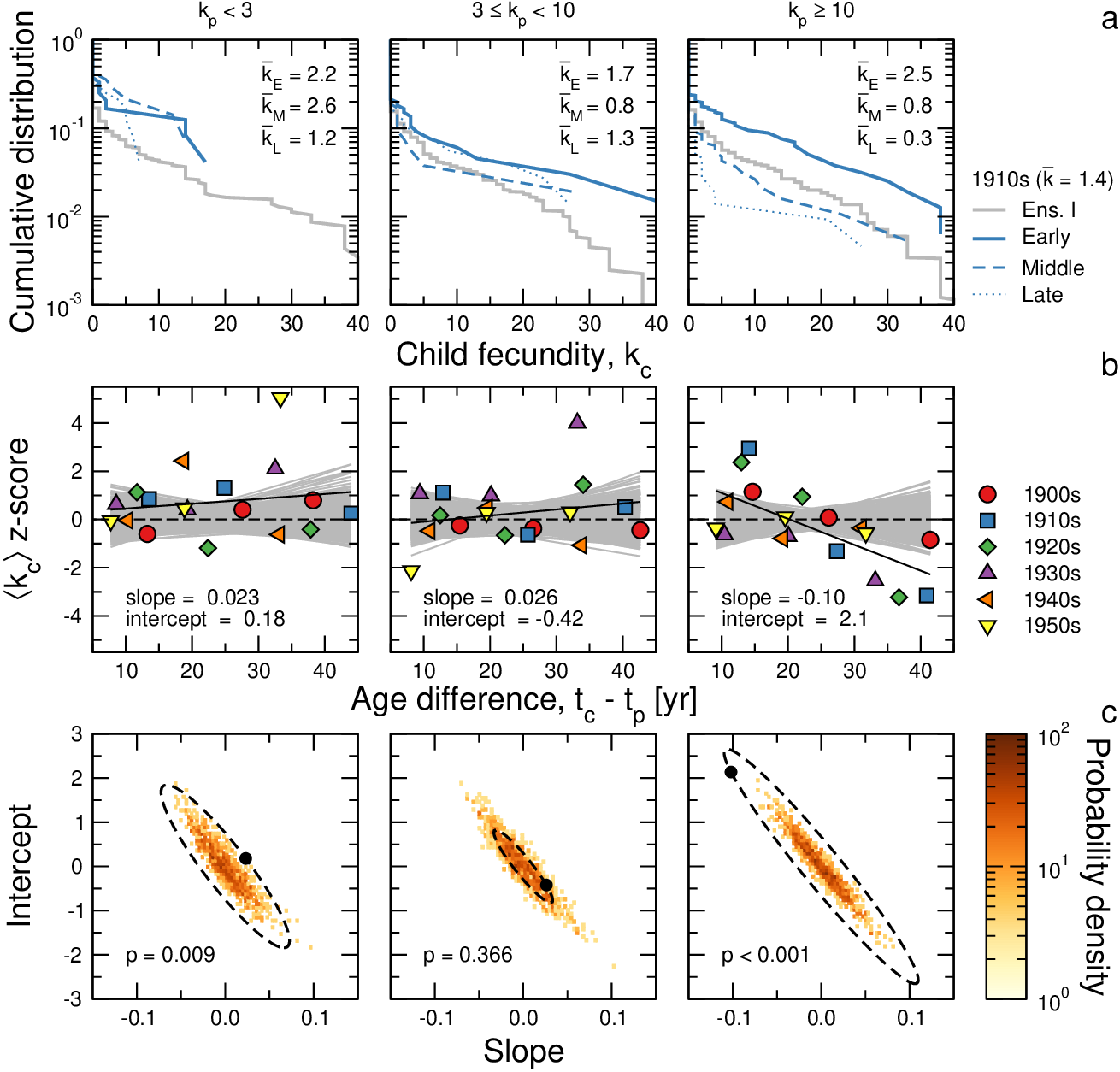}}
\caption{\figDcaption}
\label{fig:kc_t0c-t0p}
\end{figure}


\end{document}